\documentclass[%
 reprint,
 amsmath,amssymb,
 aps,
]{revtex4-1}

\usepackage{graphicx}
\usepackage{dcolumn}
\usepackage{bm}
\usepackage{float}

\begin{document}


\title{ Mechanisms for tuning clustering and degree-correlations in directed networks}

\author{G. Kashyap}
\author{G. Ambika }%
\email{e-mail: g.ambika@iiserpune.ac.in}
\affiliation{%
 Indian Institute of Science Education and Research, Pune, India - 411008
}%




\date{\today}

\begin{abstract}
With complex networks emerging as an effective tool to tackle multidisciplinary problems, models of network generation have gained an importance of their own. These models allow us to extensively analyze the data obtained from real-world networks, study their relevance and corroborate theoretical results. In this work, we introduce methods, based on degree preserving rewiring, that can be used to tune the clustering and degree-correlations in directed networks with random and scale-free topologies. They provide null-models to investigate the role of the mentioned properties along with their strengths and limitations. We find that in the case of clustering, structural relationships, that are independent of topology and rewiring schemes are revealed, while in the case of degree-correlations, the network topology is found to play an important role in the working of the mechanisms. We also study the effects of link-density on the efficiency of these rewiring mechanisms and find that in the case of clustering, the topology of the network plays an important role in determining how link-density affects the rewiring process, while in the case of degree-correlations, the link-density and topology, play no role for sufficiently large number of rewiring steps. Besides the intended purpose of tuning network properties, the proposed mechanisms can also be used as a tool to reveal structural relationships and topological constraints.


\end{abstract}

\maketitle


\section{\label{sec:level1}Introduction}

Large-scale global connectivity has become an indispensable part of our lives and as a consequence, complex networks have gained an enormous importance in multiple fields of research and applications \cite{newman2002structure,strogatz2001exploring,albert2002statistical,boccaletti2006complex}. With rapid advances in technology, abundant data about these systems has been made available and this is redefining how we perceive and understand complex networks. Analyses of these datasets have revealed a multitude of properties associated with the respective networks. Though some of these properties could be causal or consequential, some structural and functional and others independent or interrelated, it is clear that our understanding of complex networks depends crucially on our understanding of their associated properties, origins and relationships. Although, in a lot of cases, we lack proper understanding of the origins of these properties, we are nevertheless interested in studying how they affect each other and the overall structure and functioning of the network. To this end, models of network generation, both growth mechanisms \cite{barabasi1999emergence,barabasi1999mean,dorogovtsev2000structure,price1976general,bollobas2003directed,krapivsky2001organization,krapivsky2000connectivity,ostroumova2013generalized} and static mechanisms \cite{erdds1959random,watts1998collective,bollobas1998random,kim2012constructing}, have become an integral aspect of the study of complex networks.

In this regard, considerable work has been done on generating scale-free (SF) networks, using both growth and static methods. Further properties, like degree-correlations and clustering have also been incorporated into static methods \cite{newman2003mixing,newman2002assortative,noldus2015assortativity,newman2009random,park2005solution,newman2003properties} and growth models \cite{szabo2004clustering,klemm2002highly,vazquez2003growing,herrera2011generating,krot2015local}. Despite the availability of extensive literature, we find that the bulk of it is limited mostly to undirected networks. In spite of their abundant occurrence and rich structural variety, directed networks have been given significantly less attention \cite{fagiolo2007clustering,williams2014degree,foster2010edge,mayo2014mixed,roberts2012unbiased} or have been approximated to the undirected case.

In this work, we focus our attention on directed networks, with 2 important properties: 2-node degree-correlations and clustering. We use existing models to generate directed SF and random networks. We then propose Degree Preserving Rewiring (DPR) mechanisms to introduce and tune the properties of interest in these networks. Methods based on DPR allow us to isolate the role of the degree distribution and related intrinsic structural properties and focus on the aspects that are unique to the network under consideration \cite{roberts2012unbiased}\cite{van2010influence}\cite{zhou2007structural}. We compare and contrast the performance of these mechanisms on the 2 types of networks and investigate their effects and side-effects. We also study the effect of link density on the performance of these techniques.

\section{\label{sec:level2}DPR methods for tuning of clustering}

In networks, the tendency for nodes to organize into well-knit neighborhoods or form small cliques, is referred to as clustering, and is quantified by the Mean Clustering Coefficient (MCC). These substructures, occur spontaneously in most networks and play an important role in the spreading of epidemics in large communities and also effect network flow in local neighborhoods.

The MCC of a network is defined as the average of Local Clustering Coefficients (LCCs) of all nodes in the network. The LCC of a node is the ratio of number of pairs of connected neighbors of the node to the number of pairs of neighbors of the node. In other words, it measures the extent to which the neighbors of a node form closed triplets with the node of interest.

\begin{figure}[H]
\centering
\includegraphics[width=\columnwidth]{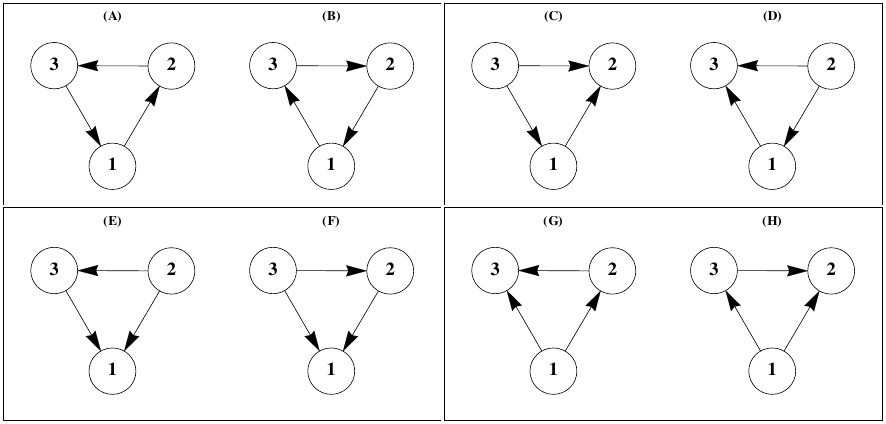}
\caption{Different types of simple closed triplets possible in a directed network are shown w.r.t node-1: (A,B)-Cycles (C,D)-Mids (E,F)-Intri and (G,H)-Outri.}
\label{fig:f1}
\end{figure}

In directed networks, as a consequence of the directionality of edges, 4 different types of simple closed triplets are possible \cite{fagiolo2007clustering}. The 4 types of triangles, namely, Cycles, Middleman-triangles (Mids), In-triangles (Intri) and Out-triangles (Outri) are shown in fig.\ref{fig:f1}. 

Given a network of size N, represented by an adjacency matrix A, and where $d_i^{in}$, $d_i^{out}$ and $d_i^\leftrightarrow$ are the in-degree, out-degree and bidirectional edges, respectively, of node i, the MCCs w.r.t each of the triangles can be calculated as follows:

\begin{subequations}
\label{eqn:1}

\begin{equation}\label{eqn:1a}
\overline{C}_{cyc}=\frac{1}{N}\sum_{i=1}^{N}\frac{(AAA)_{ii}}{d_i^{in}d_i^{out}-d_i^{\leftrightarrow}}
\end{equation}

\begin{equation}\label{eqn:1b}
\overline{C}_{mid}=\frac{1}{N}\sum_{i=1}^{N}\frac{(AA^TA)_{ii}}{d_i^{in}d_i^{out}-d_i^{\leftrightarrow}}
\end{equation}

\begin{equation}\label{eqn:1c}
\overline{C}_{int}=\frac{1}{N}\sum_{i=1}^{N}\frac{(A^TAA)_{ii}}{d_i^{in}(d_i^{in}-1)}
\end{equation}

\begin{equation}\label{eqn:1d}
\overline{C}_{out}=\frac{1}{N}\sum_{i=1}^{N}\frac{(AAA^T)_{ii}}{d_i^{out}(d_i^{out}-1)}
\end{equation}

\end{subequations}


To improve the amount of clustering in the network, we employ mechanisms that identify a suitable chain of connected nodes, that can then be reconfigured to give a closed triplet of the desired form, while preserving the degrees of the nodes involved. While the mechanisms are distinct for different types of triplets, they only differ very little from each other based on the triplet of interest.

To improve the amount of clustering w.r.t cycles, we use the following procedure:

Step-1: Calculate the initial value of $\overline{C}_{cyc}$ using (\ref{eqn:1a}). Select a node i, whose neighborhood can be suitably modified. While there is no fixed manner in which i can be chosen, the only condition that needs to be satisfied is that i must have at least 1 incoming link and 1 outgoing link. In other words, the in and out degrees of i must be greater than or equal to 1. It is also worth mentioning that i can be chosen either uniformly randomly or in a weighted manner, where the weights could be LCC of i, in-degree of i or out-degree of i. Although a weighted selection of i would seem to be more effective, we find that, for a given number of rewiring steps, the networks undergo approximately the same amount of change in clustering. Therefore, for the purpose of this work, i is chosen uniformly randomly.

Step-2: From the list of incoming neighbors of i, we randomly select a node j, and from the list of outgoing neighbors of i, we select k.

Step-3: From the list of incoming neighbors of j, we randomly select a node m, and from among the outgoing neighbors of k, we select n. If it is not possible to make either of the selections, we go back to step-1.

Step-4: If nodes m, n and i are distinct, and edges (m,n) and (k,j) do not exist, then edges (m,j) and (k,n) are rewired to (k,j) and (m,n). In this process, a cyclic triplet (j,i,k) is created.

Step-5: Calculate $\overline{C}_{cyc}$ of the network. If there is an overall increase from the initial value, then retain the change and go back to step-1.  If there is a decrease in clustering, then reject the rewiring and go back to step-1.

The above process is iterated for a predetermined number of steps or until a predetermined value of clustering coefficient is reached.

\begin{figure}[H]
\centering
\includegraphics[width=0.6\columnwidth]{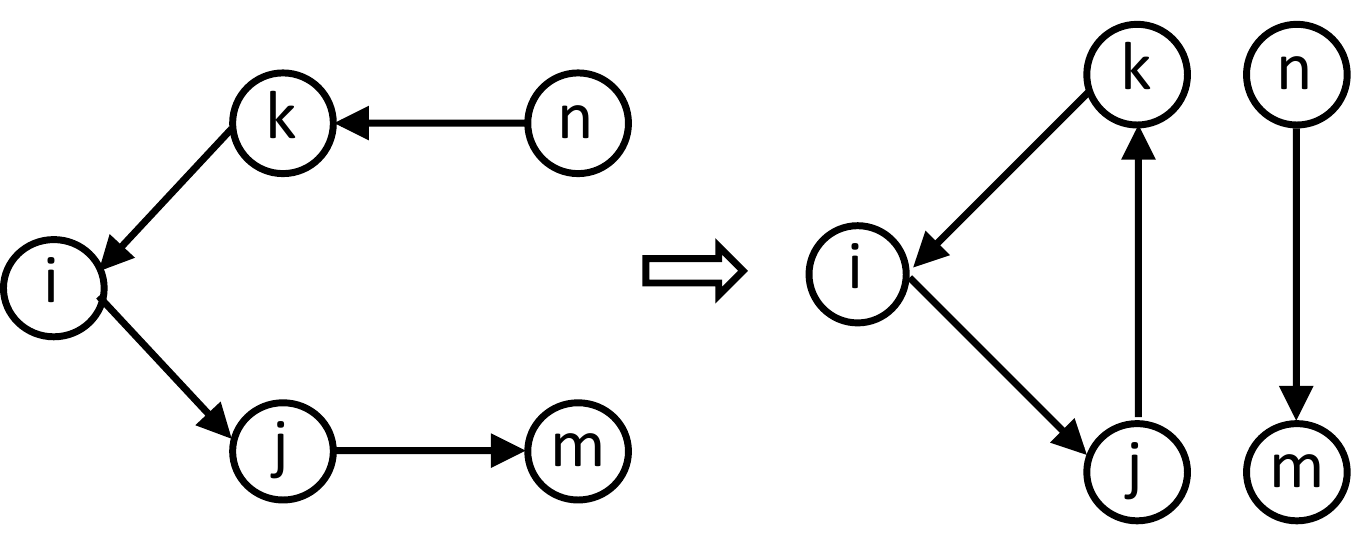}
\caption{Schematic representation of the DPR mechanism for tuning clustering w.r.t cycles.}
\label{fig:f2}
\end{figure}

To improve the amount of clustering w.r.t mids, we use the following procedure:

Step-1: Calculate the initial value of $\overline{C}_{mid}$ from (\ref{eqn:1b}). Randomly select a node i, with both in and out-degrees greater than or equal to 1.

Step-2: Select node j from among the incoming neighbors of i and node k from among the outgoing neighbors of i.

Step-3: Further, select node m from among the outgoing neighbors of j and node n from the incoming neighbors of k. If any one of them is not possible, then go back to step-1.

Step-4: If m, n and i are distinct and edges (j,k) and (n,m) do not exist, then edges (j,m) and (n,k) are rewired to (j,k) and (n,m), forming a middleman triangle (j,i,k).

Step-5: Calculate $\overline{C}_{mid}$ of the network. If there is an overall decrease in clustering, then reject the rewiring and go back to step-1, and in case of an overall increase, retain the rewiring and go to step-1.

The above process is iterated for a predetermined number of steps or until a predetermined value of clustering is reached.

\begin{figure}[H]
\centering
\includegraphics[width=0.6\columnwidth]{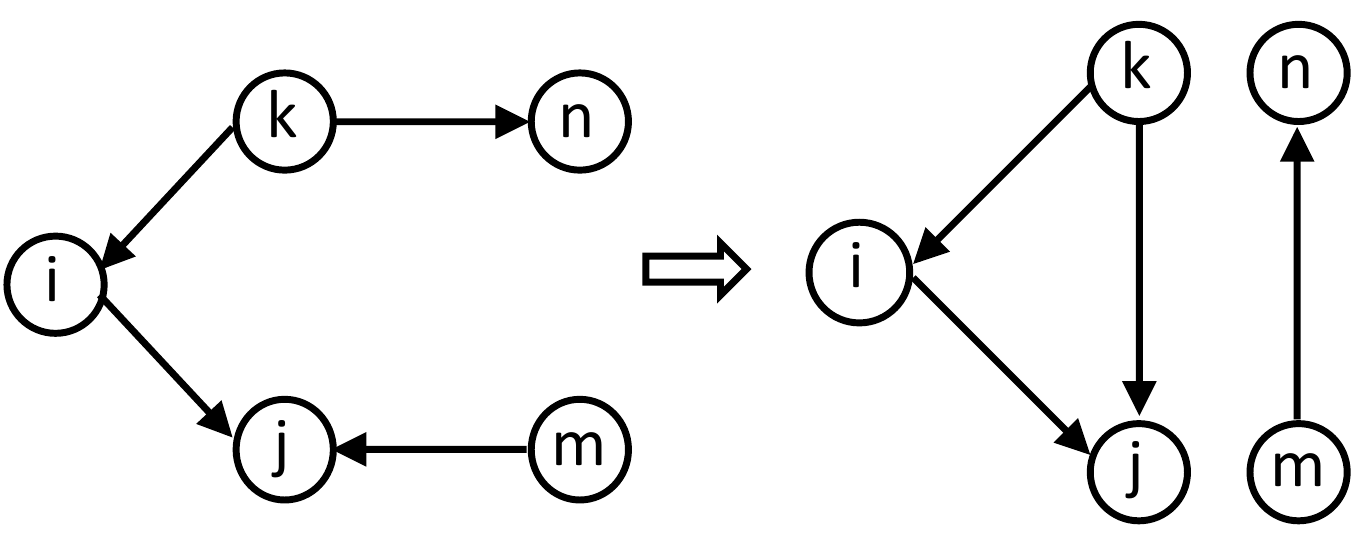}
\caption{Schematic representation of the DPR mechanism for tuning clustering w.r.t mids.}
\label{fig:f3}
\end{figure}

To improve the amount of clustering w.r.t intri, we use the following procedure:

Step-1: Calculate the initial value of $\overline{C}_{int}$ using (\ref{eqn:1c}). Randomly select a node i, with in-degree greater than 1.

Step-2: Select nodes j and k from among the incoming neighbors of i.

Step-3: Further, select node m from the neighbors (in and out) of j and node n from the neighbors (in and out) of k. If any one of them is not possible, then go back to step-1.

Step-4: If m, n and i are distinct and edges (j,k) and (n,m) do not exist and (j,m) and (n,k) exist, then edges (j,m) and (n,k) are rewired to (j,k) and (n,m), to form an in-triangle (j,i,k). Else, if nodes m, n and i are distinct, and edges (m,n) and (k,j) do not exist and (m,j) and (k,n) exist, then edges (m,j) and (k,n) are rewired to (k,j) and (m,n), once again forming an in-triangle (j,i,k).

Step-5: Calculate $\overline{C}_{int}$ of the network. If there is an overall decrease in clustering, then reject the rewiring and got back to step-1, and in case of an overall increase, retain the rewiring and go to step-1.

The above process is iterated for a predetermined number of steps or until a predetermined value of clustering is reached.

\begin{figure}[H]
\centering
\includegraphics[width=0.6\columnwidth]{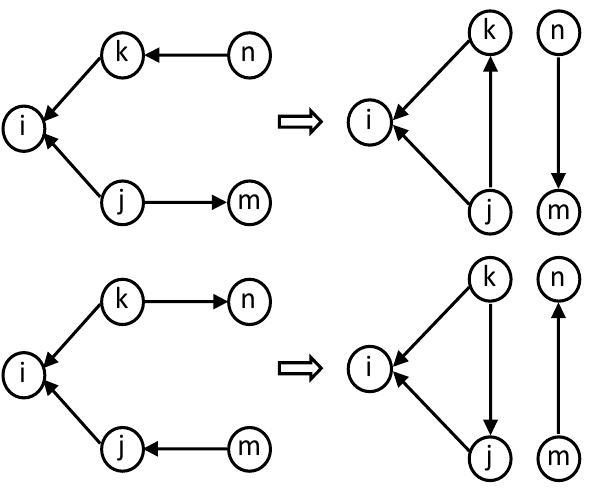}
\caption{Schematic representation of DPR mechanisms for tuning clustering w.r.t intri.}
\label{fig:f4}
\end{figure}

To improve the amount of clustering w.r.t outri, we use the following procedure:

Step-1: Calculate the initial value of $\overline{C}_{out}$ using (\ref{eqn:1d}). Randomly select a node i, with out-degree greater than 1.

Step-2: Select nodes j and k from among the outgoing neighbors of i.

Step-3: Further, select node m from the neighbors (in and out) of j and node n from the neighbors (in and out) of k. If any one of them is not possible, then go back to step-1.

Step-4: If m, n and i are distinct and edges (j,k) and (n,m) don't exist and (j,m) and (n,k) exist, then edges (j,m) and (n,k) are rewired to (j, k) and (n, m), to form an out-triangle (j,i,k). Else, if nodes m, n and i are distinct, and edges (m, n) and (k, j) do not exist and (m,j) and (k,n) exist, then edges (m,j) and (k,n) are rewired to (k,j) and (m,n), once again forming an out-triangle (j,i,k).

Step-5: Calculate $\overline{C}_{out}$ of the network. If there is an overall decrease in clustering, then reject the rewiring and go back to step-1, and in case of an overall increase, retain the rewiring and go to step-1.

The above process is iterated for a predetermined number of steps or until a predetermined value of clustering is reached.

\begin{figure}[H]
\centering
\includegraphics[width=0.6\columnwidth]{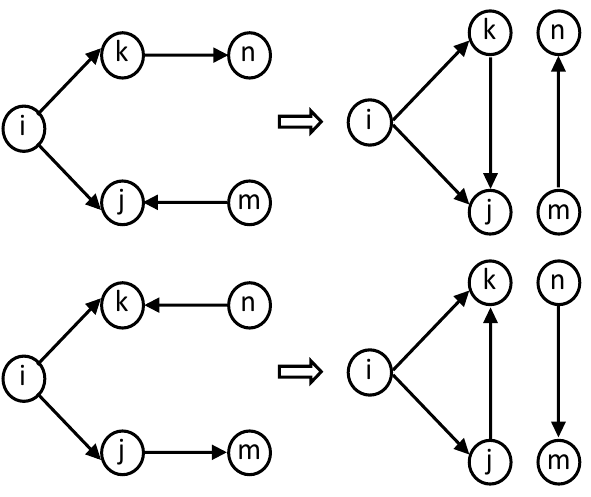}
\caption{Schematic representation of DPR mechanisms for tuning clustering w.r.t outri.}
\label{fig:f5}
\end{figure}

\section{\label{sec:level12}Results for tuning of clustering}

We test the performance of the proposed mechanisms, presented in the previous section, on a directed SF network and a directed random network and study the effectiveness of these mechanisms in generating the properties of interest. We use the directed configuration model \cite{kim2012constructing} to generate a random network, while the SF network is generated using the model given by Bollobas et al. in \cite{bollobas2003directed}. The working of the mechanisms are first studied for a given parameter value of the networks and then a further investigation is conducted for different parameter values of the respective networks. In the SF network model, the link-density($\bar{k}$) is parametrized by $\beta$. We see from \cite{bollobas2003directed} that in the N$\rightarrow\infty$ limit, $\bar{k}$ = $\beta$/(1-$\beta$). This allows us to compare performances of the rewiring mechanisms in SF networks and ER networks of approximately the same link-density.

\begin{figure}[H]
\centering
\includegraphics[width=\columnwidth]{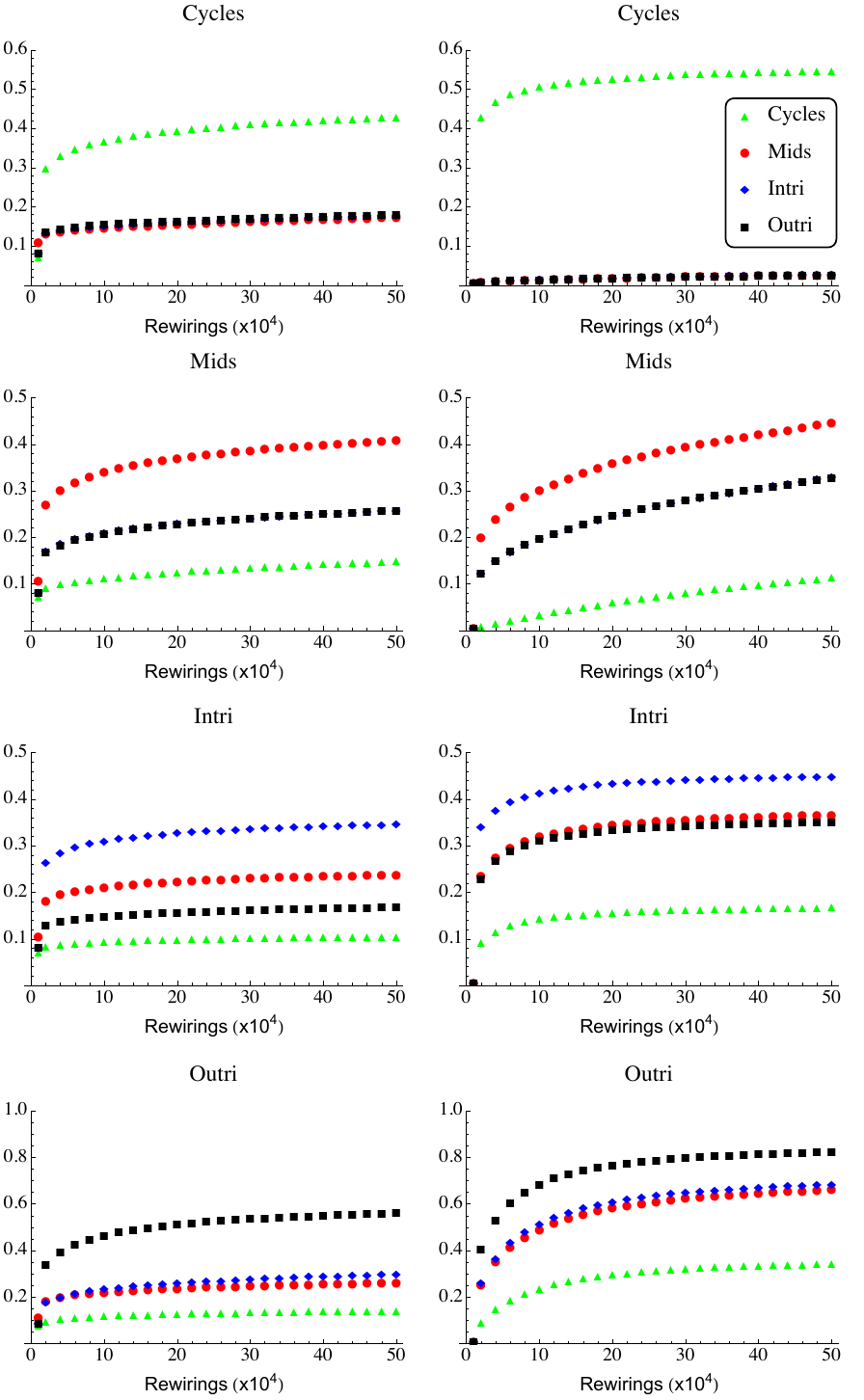}
\caption{(Color Online) MCC values for the different types of clustering generated in ER networks (right) and SF networks (left) as a function of rewiring steps. The ER networks are generated with N = $10^3$ and $\bar{k}$ = $\bar{k}^{in}$ = $\bar{k}^{out}$ = 3 and the SF networks are generated with N = $10^3$ and $\beta$ = 0.8.}
\label{fig:f6}
\end{figure}

To study the generation of any type of triangles, we generate 2 ensembles, each consisting of 100 networks. The first ensemble contains directed ER (random) networks of size N = $10^3$ and link-density equal to 5 and the second ensemble contains SF networks of size N = $10^3$ and $\beta$ = 0.8. The DPR method for the clustering of interest is applied iteratively to these ensembles of networks and the average results are shown in fig.\ref{fig:f6}.

From fig.\ref{fig:f6}, we observe that for a given N and approximately the same average degree, the mechanisms affect a substantially greater change in the ER network compared to the SF network, for the same number of rewiring steps. This is particularly pronounced in the case of $\overline{C}_{out}$, where DPR introduces twice the amount of change in ER networks than in the SF networks. The MCCs show a rapid initial increase following which they remain more or less constant or show a very slow increase.

We also notice how certain structural side-effects of the DPR mechanisms manifest themselves, independent of the network topology. For example, when a network is rewired to increase $\overline{C}_{out}$, there is noticeable increase in $\overline{C}_{int}$ and $\overline{C}_{mid}$. This can be explained by analyzing fig.\ref{fig:f5}. When the node i is selected and the links between its first and second neighbors are rewired, as a consequence of the rewiring, 2 complementary triangles, w.r.t its first neighbors, are also created. We also notice that there is almost equal increase in both $\overline{C}_{int}$ and $\overline{C}_{mid}$. A similar effect is observed in the case of $\overline{C}_{mid}$, where $\overline{C}_{int}$ and $\overline{C}_{out}$ show an increase. In the case of $\overline{C}_{int}$, there is an increase in both $\overline{C}_{out}$ and $\overline{C}_{mid}$, but unlike in the former 2 cases, the amount of change in MCC w.r.t the complementary triangles is not equal. There is greater increase in $\overline{C}_{mid}$ than $\overline{C}_{out}$. Also, this is observed only in the case of SF networks. However, none of this is observed in the case of $\overline{C}_{cyc}$ as the complementary triangles formed are also cycles.

\begin{figure}[H]
\centering
\includegraphics[width=\columnwidth]{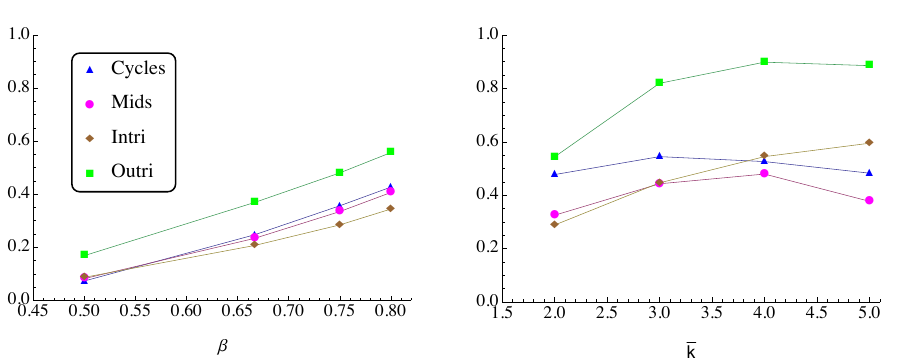}
\caption{(Color Online) Different MCC values generated in ER networks (right) and SF networks (left) for multiple parameter values. ER networks are studied for $\bar{k}$ = 2, 3, 4, 5 and SF networks for $\beta$ = 0.5, 0.66, 0.75, 0.8. The values shown are obtained after $10^6$ rewiring steps.}
\label{fig:f7}
\end{figure}

From fig.\ref{fig:f7}, we see that in SF networks, different MCCs show the same qualitative behavior but different quantitative behaviors. They get rewired to different values for the same number of rewiring steps. This requires further investigation to ascertain if it is a structural limitation. $\overline{C}_{out}$ reaches a value of ~0.6 for $\beta$ = 0.8 while $\overline{C}_{int}$ reaches only as high as 0.3 and $\overline{C}_{mid}$ and $\overline{C}_{cyc}$ go higher upto ~0.4. The behavior of all 4 MCCs is consistent with increasing values MCCs for gradually increasing values of $\beta$. In the ER network, the final values of MCCs are not as smoothly varying as in the SF network. $\overline{C}_{cyc}$ and $\overline{C}_{mid}$ have values clustered close together for $\bar{k}$ = 2, 3, 4, 5 while the values for $\overline{C}_{int}$ and $\overline{C}_{out}$ are more smoothly distributed. Overall, on comparing the results in both the network types, we find that all the DPR mechanisms for clustering are affected to some extent by the network topology.

\section{\label{sec:level3}DPR methods for tuning of correlations}

Degree-correlations measure the tendency of nodes to connect with other nodes with similar or dissimilar degrees. While it is not completely understood whether these correlations are the cause or consequence of other properties/processes, the importance of their effect on network structure and dynamics cannot be ignored. While correlations can be studied w.r.t any enumerative property (color, race etc.) or scalar property (age, income etc.) of the nodes, correlations between node-degrees gain further importance because of the interplay between the two structural properties involved.

In directed networks, since each node has an in-degree and an out-degree, 4 types of 2-node degree-correlations can be defined, namely, In-In, In-Out, Out-In and Out-Out degree correlations. Further in this paper, we will refer to a particular type of correlation as p-q, where p,q $\in$ $\{$In,Out$\}$ and p and q are associated with the source and target nodes respectively. The traditional metric used to quantify degree correlations is the Pearson correlation coefficient, $r_q^p$, as introduced in \cite{newman2003mixing}.



But, as was shown in \cite{dorogovtsev2010zero}\cite{litvak2013uncovering}\cite{van2015degree}, $r_q^p$ scales with the network size N and therefore, in the case of distributions with heavy tails, it converges to a non-negative number as N$\rightarrow\infty$. As a result, it does not lend itself well to capture the dissortativity in large SF networks. For this reason, in this work, we turn to the Spearman’s rank correlation(\ref{eqn:2a}), to quantify these dependencies. 

\begin{equation}\label{eqn:2a}
\rho_q^p=\frac{12\sum_e R_e^pR_e^q-3M(M+1)^2}{M^3-M}
\end{equation} 

where $R_e^p$ and $R_e^q$ are the ranks of source and target nodes associated with edge e, based on their p and q-degrees, respectively, and $\rho_q^p$ is referred to as the Spearman's Rho \cite{van2015degree}. Both $\rho_q^p$ and $r_q^p$ are bound in the range [-1,1], with the values being positive when nodes of similar degrees are connected by an edge and negative when nodes of dissimilar degrees are connected. The value becomes 0 when there is no net bias.


In order to design a mechanism to tune degree correlations, we start with the rewiring rule from \cite{van2010influence} and adapt it to the case of directed edges, where the identities of the source and target nodes need to be additionally preserved. We modify the procedure in \cite{van2010influence} in the following manner.

\begin{figure}[h]
\centering
\includegraphics[width=0.6\columnwidth]{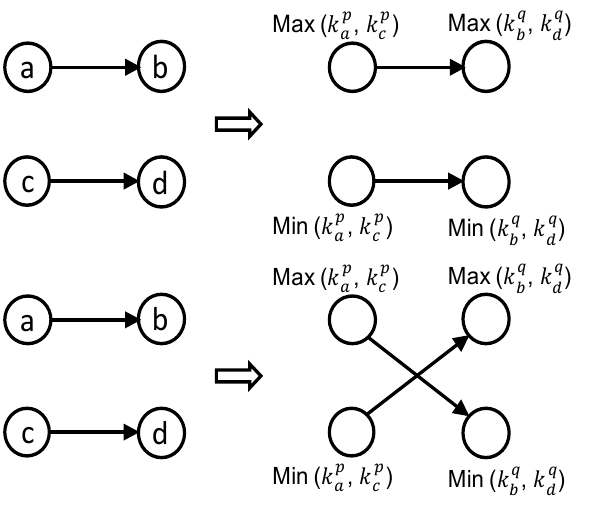}
\caption{Schematic representation of DPR mechanisms for assortative rewiring (top) and dissortative rewiring (bottom).}
\label{fig:f8}
\end{figure}

Given p,q $\in$ $\{$in, out$\}$, we randomly choose 2 links, (a,b) and (c,d). From among the 2 source nodes, a and c, we select the node with higher p-degree and from the target nodes, b and d, we select the node with higher q-degree. If they aren't identical or already connected by a link, then the existing links are deleted and new links are placed between the 2 selected nodes and the 2 remaining nodes. This is done to prevent the appearance of multi-edges and self-edges during the course of rewiring. Repeated iteration of the rewiring step (fig.\ref{fig:f8} (top)) generates a network that is assortative in p-q type of correlations.

To generate a network with dissortative p-q correlations, 2 links (a,b) and (c,d) are randomly chosen. From among the source nodes, the node with higher p-degree is selected and from the target nodes, the node with smaller q-degree is selected. Existing links are deleted and new links are placed between the 2 selected nodes and the 2 remaining nodes, while simultaneously ensuring that no multi-edges or self-loops are created. This rewiring step (Fig.\ref{fig:f8} (bottom)) is iterated over to incrementally increase the dissortativity in the network.

\section{\label{sec:level22}Results for tuning of correlations}

\begin{figure}[H]
\centering
\includegraphics[width=\columnwidth]{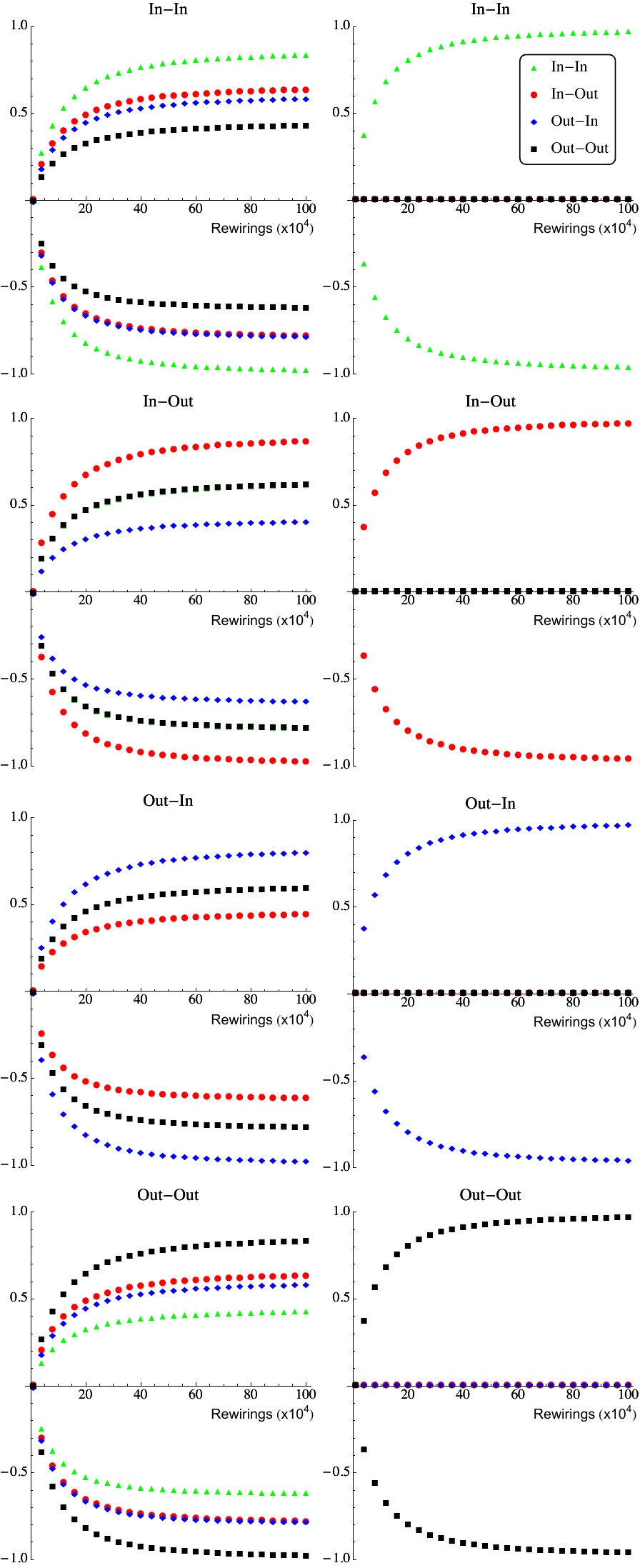}
\caption{(Color Online) Results of DPR mechanisms, associated with the 4 different types of degree-correlations, in ER networks(right) and SF networks(left). The results for assortative and dissortative rewirings, for each variant of correlation are plotted together for ease of comparison.}
\label{fig:f9}
\end{figure}

To study the generation of degree-correlations, we generate 2 ensembles, each containing 100 networks of size N = $10^4$. The SF networks are generated with $\beta$ = 0.8 and the ER networks with $\bar{k}$ = 5. The relevant assortative and dissortative rewiring mechanisms are iterated over on these ensembles and the results, that are ensemble averages, are shown in fig.\ref{fig:f9}.

\begin{figure}[H]
\centering
\includegraphics[width=\columnwidth]{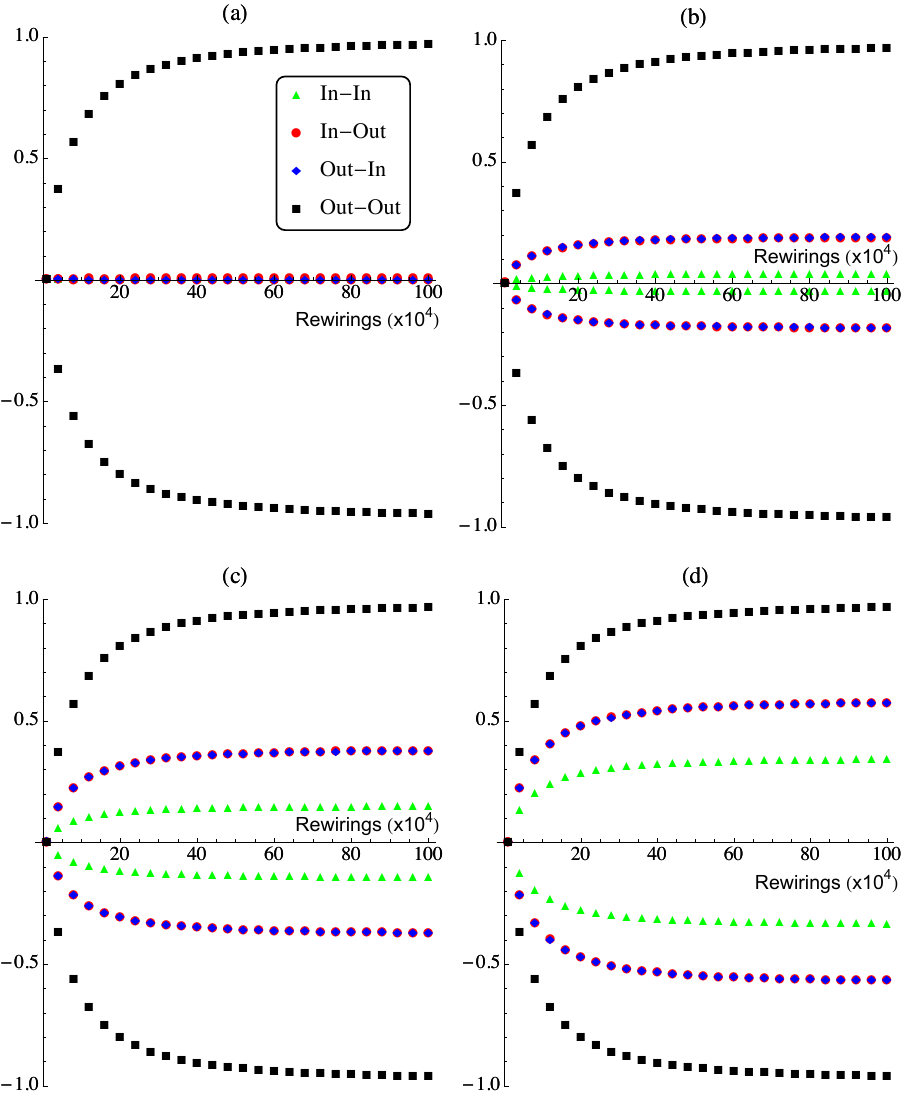}
\caption{(Color Online) Results of rewiring mechanisms for 2-node Out-Out correlations in ER networks as 1-node in-out correlation is gradually increased to take values of (a)-0, (b)-0.2, (c)-0.4 and (d)-0.6. The results extend without loss of generality to other variants of 2-node correlations.}
\label{fig:f10}
\end{figure}

In ER networks, all variants of the mechanisms work perfectly and the networks show only the relevant p-q correlations (p,q $\in$ $\{$in,out$\}$) that correspond to the respective rewiring mechanism and nothing else (fig.\ref{fig:f9} (right)). However, the SF networks present a more interesting scenario (fig.\ref{fig:f9} (left)). Although the p-q correlation of interest is introduced to the largest extent, other variants also arise with considerable magnitude. Since the same mechanisms do not result in a similar behavior in ER networks, it is evident that this behavior is not the result of rewiring itself but some other topological property. On examining the 2 types of networks for topological differences besides the distribution of degrees, we find them to be identical in all aspects except the case of 1-node correlation. We find that, in ER networks, there is no correlation between the in and out degrees of a given node, while in the case of SF networks, there is a very strong correlation. This can be traced back as an artifact of the construction process in \cite{bollobas2003directed}, where nodes appearing early in the growth process, have higher in and out degrees. To confirm 1-node correlations as the cause for the observed behavior, we introduce 1-node in-out correlations in ER networks in a systematic manner. Fig.\ref{fig:f10} shows the change in behavior of 2-node Out-Out correlation when the 1-node correlation is gradually increased.

Also, for the same number of rewiring steps, the absolute value of $\rho_q^p$ is smaller in the case of assortative rewiring and higher in the dissortative case. Another important feature is that, in the case of in-out and out-in correlations, the results for assortative and dissortative rewiring show qualitatively similar behavior with the 4 types of correlations showing 3 distinct values. This symmetric behavior is not seen in the case of in-in and out-out correlations, where the correlations taking the intermediate values further split up, so that the 4 variants take 4 distinct values. Further, this splitting only occurs in the case of assortative rewiring and not in the dissortative case, leading to further questions about 2-node and 1-node structural relationships.

\begin{figure}[h]
\centering
\includegraphics[width=\columnwidth]{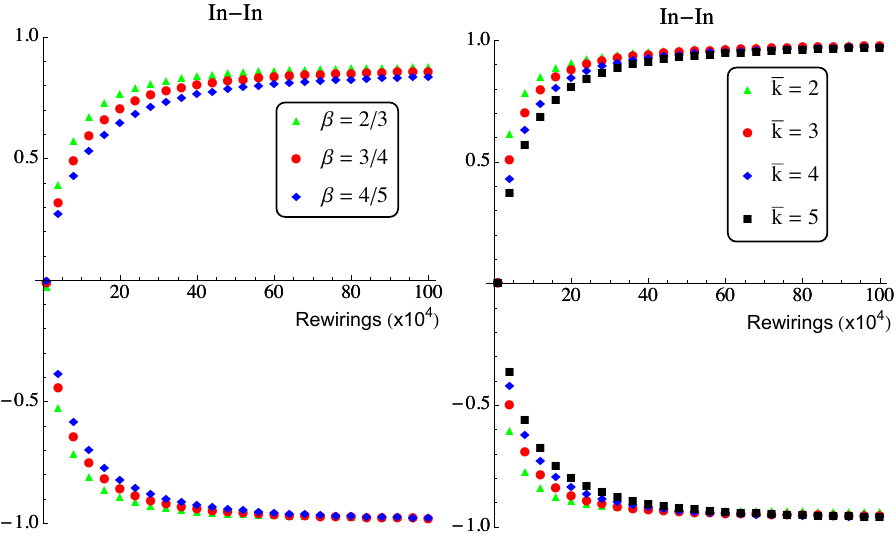}
\caption{(Color Online) Effects of link-density on the performance of rewiring mechanisms for In-In correlations in ER networks (right) and SF networks (left). Rewiring for In-Out, Out-In and Out-Out correlations also show similar qualitative and quantitative behaviors.}
\label{fig:f11}
\end{figure}

In ER networks, assortative and dissortative rewirings are studied for $\bar{k}$ = 2, 3, 4, 5. The behavior is identical in both cases, with higher link-densities showing slower rate of change (fig.\ref{fig:f11} (right)). We also see that for sufficiently large number of rewiring steps, the absolute values of $\rho_q^p$, corresponding to all $\bar{k}$, converge to a large value close to 1. The results are similar in the case of dissortative rewiring in SF networks, for corresponding values of $\beta$ (fig.\ref{fig:f11} (left)). In assortative rewiring, however, values of $\rho_q^p$ do not quite converge for increasing values of $\beta$. The network reaches lower values of $\rho_q^p$ for higher values of $\beta$, for the same number of rewiring steps.

\section{\label{sec:level4}Conclusions}

To summarize, we have presented DPR mechanisms to tune the amount of degree-correlations and clustering in directed networks with random and SF topologies. These mechanisms allow us to explore the relevant properties, independent of the topology, in our attempt to understand their role in the structural organization and functioning of the networks. They provide alternate ways to introduce and tune properties, especially when growth mechanisms fail, due to our lack of knowledge about the processes or mechanisms leading to these properties. Having mechanisms that can artificially tune the amount of clustering makes it easier to study the role of clustering in information dissemination in computer or social networks and rumor-spreading in online or offline communities. It also helps in designing and testing efficient mitigation strategies to contain epidemics in contact networks. In this regard, the ability to tune 2-node degree-correlations also plays an important role. Tuning correlations also helps us to study their effect on the robustness of networks under attack and during failure.

We find that, for both correlations and clustering, the density of links in the network does not affect any qualitative change in the working of the mechanisms. The general observation is that higher link-densities slow down the effects of rewiring. We also conclude that the topology itself affects the qualitative behavior of the mechanisms. There exist structural side-effects that are inherent in the definitions of the properties, as a result of which some properties cannot be manipulated in isolation. Basically, the results emphasize on the structural relationships present in directed networks, particularly in SF networks. Finally, although we set out to find mechanisms to tune clustering and degree-correlations in directed networks, we find that the same mechanisms double up as tools to explore further structural relationships in the networks.



%

\end{document}